Anisotropic physical properties and large critical current density in $KCa_2Fe_4As_4F_2$ single crystal

Sunseng Pyon[1], Yuto Kobayashi[1], Ayumu Takahashi[1], Wenjie Li[1], Teng Wang[2], Gang Mu[2], Ataru Ichinose[3], Tadashi Kambara[4], Atsushi Yoshida[4], Tsuyoshi Tamegai[1]

1 Department of Applied Physics, The University of Tokyo, 7-3-1 Hongo, Bunkyo-ku, Tokyo 113-8656, Japan
2 State Key Laboratory of Functional Materials for Informatics, Shanghai Institute of Microsystem and Information Technology, Chinese Academy of Sciences, Shanghai 200050, China
3 Central Research Institute of Electric Power Industry, Electric Power Engineering Research Laboratory, 2-6-1, Nagasaka, Yokosuka-shi, Kanagawa 240-0196, Japan
4 Nishina Center, RIKEN, 2-1 Hirosawa, Wako, Saitama 351-0198, Japan

**Abstract**

   We present a systematic study of electrical resistivity, Hall coefficient, magneto-optical imaging, magnetization, and STEM analyses of $KCa_2Fe_4As_4F_2$ single crystals. Sharp diamagnetic transition and magneto-optical imaging reveal homogeneity of single crystal and prominent Bean-like penetrations of vortices. Large anisotropy of electrical resistivity, with $\rho_c/\rho_{ab} > 100$, and semiconductor-like $\rho_c$ suggest that the electronic state is quasi two-dimensional. Hall effect measurements indicate that $KCa_2Fe_4As_4F_2$ is a multiband system with holes as main carriers. Magnetization measurements reveal significantly larger $J_c$ compared with that in other iron-based superconductors with different values of $J_c$ depending on the direction of magnetic field. Origin of these $J_c$ characteristics is discussed based on microstructural observations using STEM. In addition, further enhancement of $J_c$ in $KCa_2Fe_4As_4F_2$ for future application is demonstrated in terms of heavy-ion irradiation.

**Introduction**

The discovery of iron-based superconductors (IBSs) in 2008 [1] has prompted great interest in the field of condensed matter physics. Their multi-gap superconductivity due to their multiband electronic structures were studied. Not only their unconventional superconducting mechanism leading to high transition temperature, but also their potential for applications has been extensively studied and several types of IBSs are expected as candidates for future applications [2,3,4]. In the case of widely-studied IBSs such as 122-type (Ba,K)Fe$_2$As$_2$ [5], their electronic states are three-dimensional with weak anisotropy. It makes good a contrast to two-dimensional electronic states in cuprate superconductors. Recently, however, it is suggested that quasi two-dimensional electronic states emerge in novel IBSs such as 12442-type compounds [6-10] or some IBSs consisting of FeSe layers sandwiched by thick insulating layers [11,12]. The 12442-type IBSs, a series of bi-layered compounds $AB_2$Fe$_4$As$_4C_2$ ($A$ = K, Rb, and Cs; $B$ = Ca, Nd, Sm, Gd, Tb, Dy, and Ho; and $C$ = F and O) are recognized as the intergrowth of 1111- and 122-type IBSs, and reported to be superconductors with superconducting transition temperature $T_c$ = 28−37 K [6-10]. The 12442-type IBSs have double Fe$_2$As$_2$ conducting layers between two neighboring Ca$_2$F$_2$ insulating layers [6]. Based on the first-principle calculation, 10 bands crossing the Fermi level are predicted, which are more complicated than other IBSs, showing multi-band character [12]. Using polycrystalline samples of 12442-type IBSs, their electronic and magnetic structures have been investigated [14-16]. Furthermore, very recently, the growth of millimeter-sized high-quality single crystals of 12442-type IBSs was reported [17,18]. It is needless to say that studies on high-quality single crystals are essential to reveal intrinsic properties of these superconductors. Using these single crystals, electronic states with large anisotropy were revealed by the estimation of the upper critical field, $H_{c2}$ [17,18], and torque analyses [19]. In addition, neutron spin resonance measurements revealed quasi two-dimensional electronic behavior of KCa$_2$Fe$_4$As$_4$F$_2$, which is similar to those of highly anisotropic cuprate superconductors [20]. Not only elucidation of unconventional superconductivity but also the possibility of future application of this material is intriguing topics. $H_{c2}$ of KCa$_2$Fe$_4$As$_4$F$_2$ is much larger than that of other systems of IBSs, when the magnetic field is applied along the *ab* plane, and large anisotropy parameter $\gamma$ ($= H_{c2}^{ab}/H_{c2}^{c}$) ~8 was revealed near the superconducting transition [17,21,22]. Large $H_{c2}$ is helpful for future high-field applications. Although critical current density, $J_c$, in KCa$_2$Fe$_4$As$_4$F$_2$ has not been reported yet, recent report on $J_c$ in CsCa$_2$Fe$_4$As$_4$F$_2$ at 2 K under self-field is comparable to that in well-studied 122-type materials [18,23]. Further studies on $J_c$ in 12442-type IBSs are demanded.

In this paper, we characterize the superconducting properties of KCa$_2$Fe$_4$As$_4$F$_2$ single crystals by measuring anisotropic electrical resistivity, in-plane Hall coefficient, and anisotropic irreversible magnetization. The quality and microstructure of single crystals were also characterized by magneto-optical imaging and scanning transmission electron microscopy (STEM). The electronic structure of $KCa_2Fe_4As_4F_2$ is discussed by comparing it with that in other IBSs. In addition, we found significantly large in-plane $J_c$ in $KCa_2Fe_4As_4F_2$ compared with that in other IBSs, and unique anisotropy of $J_c$ depending on the direction of magnetic field. Further enhancement of $J_c$ in $KCa_2Fe_4As_4F_2$ for future application in terms of heavy-ion irradiation is also demonstrated.

**Experimental methods**

Single crystals of $KCa_2Fe_4As_4F_2$ were grown using the self-flux method with KAs as flux. Details of sample growth and basic physical properties have been reported in ref. [17]. The bulk magnetization for $H // c$-axis and $ab$-plane were measured to evaluate the $J_c$ by using a superconducting quantum interference device (SQUID) magnetometer (MPMS-5XL, Quantum Design). Electrical resistivity along the $ab$-plane ($\rho_{ab}$) and along the $c$-axis ($\rho_c$) were measured by the four-probe method. Anticipating relatively large anisotropy of transport properties based on the previous measurements [17], care was taken to prepare electrical contacts so that the current flows uniformly. In the case of $\rho_{ab}$ measurement, consistency of the values evaluated on the top and bottom surfaces was carefully examined. The Hall voltage was measured with a Quantum Design physical property measurement system (PPMS) with the AC transport option with current density of ~11 A/cm$^2$. The Hall voltage was obtained from the antisymmetric part of transverse voltage by subtracting the positive and negative magnetic field data. Macroscopic homogeneity of superconductivity was examined by measuring the critical state field profile using magneto-optical (MO) imaging. For MO imaging, an iron-garnet indicator film was placed in direct contact with the sample surface, and the whole assembly was attached to the cold finger of a He-flow cryostat (Microstat-HR, Oxford Instruments). MO images were acquired by using a cooled-CCD camera with 12-bit resolution (ORCA-ER, Hamamatsu). Cross-sectional observations of the single crystals were performed with a high-resolution STEM (JEOL, JEM-2100F). Enhancement of $J_c$ was attempted by irradiating 2.6 GeV U ion along the $c$-axis at RIKEN Ring Cyclotron in RI Beam Factory operated by RIKEN Nishina Center and CNS, The University of Tokyo. Columnar defects at a dose-equivalent matching fields of $B_\Phi = 40$ kG were created.

**Results and discussions**

Figure 1 shows the temperature dependence of magnetization ($H // c$-axis) at 5 Oe for $KCa_2Fe_4As_4F_2$ (510 x 390 x 7.8 μm$^3$). $T_c$ defined by the onset of diamagnetism is 33.0 K and $\Delta T_c$ is less than 1 K. The quality of the sample is almost the same as that reported in the previous publication [17].

MO imaging of $KCa_2Fe_4As_4F_2$ was performed to evaluate its homogeneity and quality. Figures 2(a) and 2(b) display MO images of $KCa_2Fe_4As_4F_2$ in the remanent state at (a) 5 K and (b) 20 K, respectively, after cycling the field up to 1.2 kOe for 0.2 s. The crystal for MO measurements are carefully cut into a parallelepiped with smooth surface and no visible cracks as shown in the optical micrograph of Fig. 2(c). At 5 K, the magnetic field is mostly shielded and only partially penetrates the sample due to large $J_c$ and the limitation of the value of the applied field. A faint horizontal white line near the center of the crystal may indicate a small crack which is invisible from the surface. At 20 K, the magnetic field reaches the center of the sample and the MO image of the right part shows the critical state field profile expected for a uniform thin-plate superconductor with clear current-discontinuity lines (*d* lines), although the magnetic field in the left half of the sample is not fully penetrated. These results indicate that the sample is fairly homogeneous with a weak variation of $J_c$ in the crystal. Local magnetic induction profiles at different temperatures taken along the broken line in Fig. 2(a) are shown in Fig. 2(d). Magnetic induction profiles at higher temperatures of 20~30 K show rooftop patterns, indicating that the large and homogeneous current flows throughout the sample. At lower temperatures, however, the magnetic induction at the center of the sample are almost zero, since $J_c$ in the sample is so high that the applied field up to 1.2 kOe is shielded. From the value of the trapped field, $J_c$ can be roughly evaluated [24]. At 20 K, trapped magnetic induction $\Delta B$ is 581 G. Using the approximate formula between $\Delta B$ and $J_c$ for a thin superconductor with a thickness $t$, $J_c \sim \Delta B /(t * \beta)$, with $t = 9$ μm ($\beta$ is a parameter determined by the sample dimensions and he distance between the sample surface and the garnet film, and $\beta \sim 3.3$ in the present case) [24], $J_c$ at 20 K under the self-field is evaluated as 0.19 MA/cm$^2$.

The superconducting transition and electrical characteristics are also evaluated by the temperature dependence of $\rho_{ab}$. Two kinds of samples, No. 1 (920 x 392 x 8 μm$^3$) and No. 2 (878 x 236 x 3 μm$^3$) with different contact configurations are measured. The contact configuration for the sample No. 1 and No. 2 are shown in Fig. 3(a). Temperature dependences of $\rho_{ab}$ of the two samples are shown in Fig 3(b). The absolute value of $\rho_{ab} \sim 300$ μΩcm at room temperature and their temperature dependences are consistent with each other. Both sample show sharp superconducting transitions at ~34.0 K. When the electrical resistivity of anisotropic crystals is evaluated, it is possible that the resistivity is estimated incorrectly due to inhomogeneous current flow, in particular along the direction perpendicular to the surface. However, the fact that $\rho_{ab}$ of sample No. 2, where voltage contacts are attached on the opposite surface to the current contacts agrees with that of sample No. 1 with all contacts on the same surface, suggest that we can ignore the possible error in the absolute value of $\rho_{ab}$ due to sample inhomogeneity and anisotropy. As shown in Fig. 3(b), $\rho_{ab}(T)$ shows a tendency of saturation with increasing temperature. Similar saturating behaviors are also observed such as in K-doped BaFe$_2$As$_2$ [25], or CaKFe$_4$As$_4$ [26,27], and values of $\rho_{ab}$ at room temperature are very similar. In general, an electrical resistivity is inversely proportional to the mean free path of electrons $l$ in metals. As Ioffe and Regel argued, metallic conduction occurs only when $l$ is larger than the interatomic spacing $a$ [28]. So maximum resistivity is limited, and that is maximized when $l$ is comparable to $a$. This limit known as Mott–Ioffe–Regel (MIR) limit in universally observed in several metals, and the values of saturated resistivity around 100-400 μΩcm were reported [29]. By considering similarity of $\rho_{ab}(T)$ and lattice parameter between $KCa_2Fe_4As_4F_2$ and K-doped $BaFe_2As_2$ or $CaKFe_4As_4$, $\rho_{ab}$ ~ 300 μΩcm at room temperature is more plausible than that ~ 1.3 mΩcm in previous report [17]. We tried to estimate the residual resistivity at 0 K, $\rho(0\ K)$, by curve fitting using the equation $\rho(T) = \rho(0\ K) + A_1T + A_2T^2$ and $\rho(T) = \rho(0\ K) + AT^n$, over the range of 40 to 80 K to evaluate the sample quality. We observed negative $\rho(0\ K)$ values obtained by these two fittings due to the relatively large $T$-linear contribution in this temperature range. Similar negative $\rho(0\ K)$ was also observed in $Ba_{1-x}K_xFe_2As_2$ [30]. The obtained exponent of $n$ ~ 1.3 for $KCa_2Fe_4AsF_2$ is a little bit smaller than that of $n$ ~ 1.4-2.0 in $Ba_{1-x}K_xFe_2As_2$, which suggests larger $T$-linear contribution. Instead of the extrapolated $\rho(0\ K)$, the sample quality was evaluated by the resistivity value just above $T_c$, $\rho(36\ K)$, which is 28.2 and 23.1 μΩcm for sample No. 1 and No.2, respectively. With these values of $\rho(36\ K)$, the residual resistivity ratio (RRR) is defined by $\rho(300\ K)/\rho(36\ K)$, resulting to RRR ~ 11.4 or 12.3, respectively. In a similar compound of $CaKFe_4As_4$, RRR ($\rho(300\ K)/\rho(35\ K)$) of ~15 has been reported [26]. Relatively large RRR, small residual resistivity, and sharp transition width indicate that our crystals are of high quality.

Figures 4(a) and 4(b) show the temperature dependence of $\rho_{ab}$ of sample No.1 sample for $H//c$-axis and $H//ab$-plane measured at various magnetic fields up to 50 kOe. As the magnetic field is increased, the superconducting transition for $H//c$-axis broadens significantly compared with that for $H//ab$-plane. Two kinds of irreversibility fields, for fields along $c$-axis ($H_{irr}^c$) or $ab$-plane ($H_{irr}^{ab}$), were also estimated from $\rho_{ab}$ data. They are defined by the criteria of $\rho_{ab}$ ~ 0.5 μΩcm. Obtained temperature dependences of $H_{irr}^c$ and $H_{irr}^{ab}$ are shown in Fig. 4(c). We also evaluate two kinds of $H_{c2}$, for fields along $c$-axis ($H_{c2}^c$) or $ab$-plane ($H_{c2}^{ab}$), by applying the Wethamer-Helfand-Hohenberg (WHH) formula, $H_{c2}(0\ K) = -0.693T_c dH_{c2}/dT(T = T_c)$ [31], and obtained $H_{c2}(T)$ as shown in Fig. 4(c). $H_{c2}$ is defined by the two different criteria of $0.9\rho_n$ and $0.5\rho_n$. Here, $\rho_n$ is the normal state resistivity estimated from the extrapolation of the resistivity using the power-law form with $n = 1.3$ as shown in the broken line in Fig. 4(a). Estimated $H_{c2}^{ab}(0\ K)$ is 10,820 and 3,012 kOe, and $H_{c2}^c(0\ K)$ is 1,326 and 407 kOe, for $0.9\rho_n$ and $0.5\rho_n$, respectively. Using these data, the anisotropy parameter $\gamma = H_{c2}^{ab}/H_{c2}^c$ is also evaluated as 8.2 and 7.4 for $0.9\rho_n$ and $0.5\rho_n$, respectively. These estimated values of $\gamma$ and $H_{c2}^{ab}(0\ K)$ are similar to that in the previous report [17], and significantly larger than those of other IBSs [26,32-34].

To discuss the anisotropy of $KCa_2Fe_4As_4F_2$, we measured $\rho_c$ for sample No. 3 (520 x 326 x 13 μm$^3$). A schematic drawing of the electrical contacts attached to $KCa_2Fe_4As_4F_2$ for $\rho_c$ measurements is shown in Fig. 5(a). Compared with $\rho_{ab}$, the magnitude and temperature dependence of $\rho_c$ are very different. As shown in Fig. 5(b), the $\rho_c$ at room temperature is 35 mΩcm, which is almost 100 times larger than the value of $\rho_{ab}$ shown in Fig. 3(b). In the simplest scenario, the anisotropy of the resistivity should be equal to $\gamma^2$. $\rho_c/\rho_{ab}$ using the $\rho_{ab}$ of sample No.1 is shown in Fig. 5(c). $\rho_c/\rho_{ab}$ at room temperature is ~100, which is close to $\gamma^2$ evaluated from the anisotropy of $H_{c2}$. $\rho_c/\rho_{ab}$ increases with decreasing temperature, reaching a very large value of ~1,800 around $T_c$. These values are significantly larger than that in doped and non-doped BaFe$_2$As$_2$ (~100 [34] or ~4 [35,36]). The value of $\rho_c/\rho_{ab}$ larger than 100 with increasing trend with decreasing temperature are also observed in cuprate superconductors [37-39] and some IBS compounds such as (Li$_{0.84}$Fe$_{0.16}$)OHFe$_{0.98}$Se [11] and Li$_x$(NH$_3$)$_y$Fe$_2$Se$_2$ [12], where highly two-dimensional electronic states are suggested. In La$_{2-x}$Sr$_x$CuO$_4$, $\rho_c/\rho_{ab}$ for over-doped with $x \sim 0.3$, where Fermi-liquid like metallic resistivity ($\rho \sim T^\alpha$, $\alpha > 1$) is observed, is almost temperature independent. On the other hand, that for under-doped with $x < 0.2$ increases significantly with decreasing temperature [37], which is similar to that in KCa$_2$Fe$_4$As$_4$F$_2$. These results can be understood by the quasi two-dimensional Fermi-surface sheets revealed by the first-principles calculations [8,9,13]. Furthermore, temperature dependence of $\rho_c$ shows a broad maximum with its maximum at around 90 K, and a tiny kink just above $T_c$. Similar behavior is reported in CsCa$_2$Fe$_4$As$_4$F$_2$ single crystal [18]. A broad maximum in $\rho_c$-$T$ has also been observed in the same 12442-type of CsCa$_2$Fe$_4$As$_4$F$_2$ [18], or (Li$_{0.84}$Fe$_{0.16}$)OHFe$_{0.98}$Se [11], and Li$_x$(NH$_3$)$_y$Fe$_2$Se$_2$ [12]. This crossover of temperature dependence of $\rho_c$ may suggest opening of pseudogap in KCa$_2$Fe$_4$As$_4$F$_2$ as suggested by ref. [40-42].

The Hall resistivity $\rho_{yx}$ as a function of magnetic field up to 50 kOe at several temperatures are shown in Fig. 6(a). Dimensions of the sample are 784×730×12 μm$^3$. In the whole range, $\rho_{yx}$ is positive and shows linear field dependence up to 50 kOe. The Hall coefficient $R_H$ in KCa$_2$Fe$_4$As$_4$F$_2$ obtained from $\rho_{yx}$ is plotted in Fig. 6(b). The sign of $R_H$ is positive in the whole temperature range, which is consistent with self hole-doping scenario based on the first-principle calculation [13]. $R_H$ increases with decreasing temperature from 300 K to 80 K, and stays nearly constant below 80 K. The absolute value of $R_H$ around $T_c$ is two times larger than that at 300 K. The absolute value of $R_H$ in our single crystal is smaller than that of $R_H$ in polycrystalline sample [6,14]. Around 75 K where $R_H$ takes its maximum value, $R_H$ of single crystal and polycrystalline sample are ~1.5 and 2.0 x 10$^{-3}$ cm$^3$/C, respectively. Larger $R_H$ in polycrystalline sample could be originated from the increase of Hall resistivity caused by lower packing density and the anisotropy of the Hall coefficient. The absolute value of $R_H$ at 300 K, ~0.7 x 10$^{-3}$ cm$^3$/C, is similar to that of other IBSs such as Ba(Fe$_{1-x}$Co$_x$)$_2$As$_2$ (~1.1 x 10$^{-3}$ cm$^3$/C for $x = 0.1$) [43], Ba$_{1-x}$K$_x$Fe$_2$As$_2$ (~0.5 x 10$^{-3}$ cm$^3$/C for $x \sim 0.5$) [44] or CaKFe$_4$As$_4$ (~0.4 x 10$^{-3}$ cm$^3$/C) [26]. It is noteworthy that doped carriers to FeAs layer in KCa$_2$Fe$_4$As$_4$F$_2$ should be almost the same as those in Ba$_{0.5}$K$_{0.5}$Fe$_2$As$_2$ and CaKFe$_4$As$_4$, since the number of doped hole per Fe is the same in these three compounds. Similar values of $R_H$ in these compounds are consistent to the simple hole-doping scenario. The temperature dependence of $R_H$ shown in Fig. 6(b) is consistent with a multiband electronic structure of KCa$_2$Fe$_4$As$_4$F$_2$. In a simple single-band system, the Hall coefficient is given by $R_H = 1/nqc$, where $q$ is the charge of a carrier, $n$ is the carrier density, and $c$ is the speed of light, and $R_H$ is $T$-independent. By contrast, the Hall coefficient in a multiband system, for instance, consisting of electron and hole bands, is given by $R_H = (n_h\mu_h^2 - n_e\mu_e^2) / [e((n_h\mu_h + n_e\mu_e)^2)]$, where $n_h$ ($n_e$) is the density of holes (electrons) and $\mu_h$ ($\mu_e$) is the mobility of holes (electrons). The Hall coefficient in a multiband system can be temperature-dependent. The obtained results indicate that $KCa_2Fe_4As_4F_2$ is a multiband system, which is consistent with the band structure calculation [13].

The $J_c$ in $KCa_2Fe_4As_4F_2$ was evaluated by measuring the irreversible magnetization. First, in-plane $J_c$ for $H//c$-axis, which we simply call $J_c$, was evaluated from magnetization measurements. In the conventional method, $J_c$ can be evaluated using the extended Bean model $J_c = 20\Delta M/a/(1-a/3b)$, where $\Delta M$ [emu/cm$^3$] is $M_{down} - M_{up}$. $M_{up}$ and $M_{down}$ are the magnetization when sweeping the field up and down, respectively, and $a$ [cm] and $b$ [cm] are the sample width and length ($a < b$) [45-47]. For some magnetization data, however, since the self-field is significant, $\Delta M$ is reduced in the return branch and causes a non-negligible errors in the calculation of $J_c$. So $J_c$ was calculated from the magnetization of the second quadrant of the magnetic hysteresis loop in Fig. 7(a) using the extended isotropic Bean model, $J_c = 40M_{down}/a/(1-a/3b)$, after subtracting linear background, as described in ref. [48]. Figure 7 (a) shows the magnetic field dependence of magnetization at various temperatures. Evaluated $J_c$ as a function of temperature is summarized in Fig. 7(b). $J_c$ at 2 K under the self-field is 8.2 MA/cm$^2$. This value of $J_c$ is significantly larger than those of other IBSs in the same condition, such as $Ba(Fe,Co)_2As_2$ (1.0 MA/cm$^2$) [49], $BaFe_2(As,P)_2$ (1.4 MA/cm$^2$) [50], $(Ba,K)Fe_2As_2$ (2.4 MA/cm$^2$) [51], and $CaKFe_4As_4$ (1.6 MA/cm$^2$) [27]. The value of $J_c$ at 20 K under the self-field is 0.38 MA/cm$^2$. This value agrees reasonably well with that evaluated from the analysis of the MO image. A slight underestimation of $J_c$ evaluated from MO image may be caused by the effect of observed defect or a slightly larger gap between the sample surface and the garnet film. $J_c$ decreases monotonically with the magnetic field without showing the peak effect in the whole temperature range. At temperatures above 20 K, the magnetic field dependence of $J_c$ becomes significant as shown in Fig. 7(b). This is consistent with the low irreversibility field near $T_c$ as described in Fig. 4(c). Next, we demonstrate further enhancement of $J_c$ by swift-particle irradiation. It is well known that point or columnar defects created by swift-particle irradiation work as pinning centers for vortices and increase $J_c$ [49]. We introduced columnar defects by using 2.6 GeV U-ion irradiation at a dose equivalent field of $B_\Phi = 40$ kG. The $J_c$ as a function of magnetic field of the $KCa_2Fe_4As_4F_2$ is summarized in Fig. 8. The $J_c$ in 2.6 GeV U-ion irradiated sample was calculated by the same method for that in the pristine sample. After the irradiation, $J_c$ at 2 K under the self-field is enhanced up to ~19 MA/cm$^2$, which is more than twice the value of the pristine crystal. The largest value of $J_c$ obtained at 2 K under self-field in 2.6 GeV U-irradiated $KCa_2Fe_4As_4F_2$ is slightly larger than those in heavy ion-irradiated $(Ba,K)Fe_2As_2$, ~15 MA/cm$^2$ [52]. Considering the fact that $J_c$ can be enhanced with adding proper defects as pinning centers by changing conditions of irradiation, further increase of $J_c$ can be expected in $KCa_2Fe_4As_4F_2$ with artificial defects.

In the case of compounds with tetragonal symmetry, three kinds of $J_c$ should be considered. One of them is in-plane $J_c$ when the field is applied along the $c$-axis as discussed above. The others are two independent critical current densities for $H//ab$-plane, one flows in the $ab$-plane and another flows along the $c$-axis. We tentatively designate the former as $J_{c2}$ and the latter as $J_{c3}$ (following the definition in ref. [27]). However, it is difficult to evaluate $J_{c2}$ and $J_{c3}$ independently without large difference between $J_{c2}$ and $J_{c3}$ as in the case of CaKFe$_4$As$_4$ [27]. Hence, we only evaluated average $J_c$ for $H//ab$-plane, $J_c^{H//ab}$, using the extended Bean model. Since the return branch cannot be estimated well due to the small magnetization compared with the background signal, $J_c$ is calculated using the conventional extended Bean model $J_c = 20\Delta M/a(1-a/3b)$. Figure 9(b) shows temperature dependence of $J_c^{H//ab}$ under self-field together with the self-field $J_c$ for $H//c$-axis. Compared with $J_c$ for $H//c$-axis, the absolute value of $J_c^{H//ab}$ is smaller, especially below 10 K. The self-field $J_c^{H//ab}$ at 2 K is ~0.9 MA/cm$^2$. Anisotropic $J_c$ has not been studied properly in other iron-based and even cuprate superconductors. For instance, Ba(Fe,Co)$_2$As$_2$ and Fe(Te,Se) single crystals are reported to show isotropic $J_c$ [49,53]. On the other hand, $J_c^{H//ab}$ in CaKFe$_4$As$_4$ shows opposite trend ($J_c^{H//ab} > J_c$ for $H//c$-axis) [54,55], and that in the same 12442-tupe of CsCa$_2$Fe$_4$As$_4$F$_2$ at 1.8 K shows similar trend ($J_c^{H//ab} < J_c$ for $H//c$-axis) [17]. The difference between $J_c$ for $H//c$-axis and $J_c^{H//ab}$ is shown in Fig. 9(b). $J_c$ under the self-field for $H//c$-axis is ~9 and ~1.5 times larger than $J_c^{H//ab}$ at 2 K and above 15 K, respectively. While temperature dependence of $J_c^{H//ab}$ below 10 K is weaker than that of $J_c$ for $H//c$-axis, values and temperature dependence of both components of $J_c$ become similar above 15 K.

One of the possible reasons for the significantly large and anisotropic $J_c$ could be the presence of natural defects in the pristine KCa$_2$Fe$_4$As$_4$F$_2$. In the case of CaKFe$_4$As$_4$, where anisotropic $J_c$ was also observed, novel planar defects nearly parallel to the $ab$-plane were detected by STEM observations [27]. Planar defects can work as pinning centers for vortices, causing enhancement of $J_{c2}$, or block electrical current flowing along the $c$-axis leading to reduction of $J_{c3}$. The former effect may be more dominant in CaKFe$_4$As$_4$. To give a hint for the anisotropic $J_c$, we performed high-resolution STEM observations. Figures 10(a)-(d) show STEM images on the cross section parallel to the $c$-axis in the pristine and 2.6 GeV U-irradiated KCa$_2$Fe$_4$As$_4$F$_2$. Clear periodic structure along the $c$-axis in Fig. 10 (b) corresponds to half unit cell of KCa$_2$Fe$_4$As$_4$F$_2$. Horizontal black line-like defects are observed as shown in yellow broken squares in Figs. 10(a), 10(c), and 10(d), respectively. These planar defects extending more than 400 nm are randomly distributed, and the separation between defects is larger than 100 nm. The lower density of defects and longer defects along the $ab$-plane in KCa$_2$Fe$_4$As$_4$F$_2$ should reduce $J_{c3}$ effectively rather than increase $J_{c2}$ as opposed to CaKFe$_4$As$_4$, leading to the net reduction of $J_c^{H//ab}$. So the observed defects are possible origin of anisotropic $J_c$, where $J_c^{H//ab}$ is lower than $J_c$ for $H//c$-axis. On the other hand, no characteristic defect structure along the $c$-axis was observed in the pristine KCa$_2$Fe$_4$As$_4$F$_2$ which may contribute to increase $J_c$ for $H//c$-axis. Instead, as shown in Fig. 10(b), regions with dark contrasts with 5-10 nm scale are distributed. They could be the origin of the large $J_c^{H//ab}$ in $KCa_2Fe_4As_4F_2$. Lattice strains or chemical inhomogeneities of some kind can be the cause of these regions with dark contrasts. On the other hand, in 2.6 GeV U-irradiated $KCa_2Fe_4As_4F_2$, clear columnar defects along the *c*-axis are observed as shown in Figs. 10(c) and 10 (d), although these defects look a little discontinuous compared with those in 2.6 GeV U-irradiated K-doped $BaFe_2As_2$ [48,56]. Possible reason for this difference is that the threshold energies of the irradiated ions to create the continuous columnar defects are different for each material. Actually, it has been confirmed that the continuity and radius of columnar defects are quite different between $Ba(Fe,Co)Fe_2As_2$ and $(Ba,K)Fe_2As_2$ irradiated by the same 2.6 GeV U ions [56]. Observed columnar defects should work as pinning center for vortices to increase in-plane $J_c$ as shown in Fig. 8. It should be noted that $J_c$ of the irradiated crystal in higher temperature and magnetic field range shown in Fig. 8 is significantly larger than that of the pristine crystal shown in Fig. 7(b). This indicates that irreversibility field above 20 K is enhanced by pinning of vortices introduced by columnar defect. Higher performance of $J_c$ around 20 K by increasing pinning force is more advantageous for the operation of them using He-free refrigeration systems in future [2]. Moreover, further enhancement of $J_c$ of $KCa_2Fe_4As_4F_2$ in a wide temperature range is expected. In the case of $Ba_{0.6}K_{0.4}Fe_2As_2$, degree of enhancement of $J_c$ by heavy-ion irradiation is controlled not only by the matching field, but also by the energy and species of ions [52]. Further enhancement of $J_c$ by controlling the irradiation conditions is demanded in future.

**Conclusion**

We have presented a systematic study of anisotropic physical properties and critical current density in $KCa_2Fe_4As_4F_2$. The sharp onset of diamagnetic shielding and the magneto-optical image reveal the homogeneity of single crystal and prominent Bean-like penetrations of vortices. Temperature dependence of $\rho_{ab}$ shows a tendency of saturation at high temperatures with a value ~300 μΩcm at room temperature. This is comparable to values of $Ba_{0.5}K_{0.5}Fe_2As_2$ and $CaKFe_4As_4$ with similar doping levels, and is consistent to the universality of MIR limit in metals. A large anisotropy of electrical resistivity with $\rho_c/\rho_{ab}$ ~ 100 at room temperature and semiconductor-like $\rho_c$ suggest quasi-two-dimensional electronic state. $R_H$ analysis indicates that $KCa_2Fe_4As_4F_2$ is a multiband system and holes are dominant carriers. The irreversible magnetization reveal significantly larger $J_c$ compared with that in other iron-based superconductors, and large anisotropy of $J_c$ depending on the direction of magnetic field. Origin of anisotropic $J_c$ may be caused by randomly distributed planar defect and scattered dark contrasts as observed in STEM images. Further enhancement of $J_c$ up to 19 MA/cm$^2$ at 2 K under self-field is also demonstrated by irradiating 2.6 GeV U at $B_\Phi$ = 40 kG. Significantly large enhancement of $J_c$ is sustained even at high temperatures and high fields.

**Acknowledgements**

This work was partially supported by a Grant in Aid for Scientific Research (A) (17H01141) from the Japan Society for the Promotion of Science (JSPS).

**References**

[1] Y. Watanabe, M. Hirano, and H. Hosono, J. Am. Chem. Soc. **130**, 3296 (2008).

[2] H. Hosono, A. Yamamoto, H. Hiramatsu, and Y. Ma, Mater. Today 21, 278 (2018).

[3] H. Huang, C. Yao, C. Dong, X. Zhang, D. Wang, Z. Cheng, J. Li, S. Awaji, H. Wen, and Y. Ma, Supercond. Sci. Technol. **31**, 015017 (2018).

[4] S. Pyon, D. Miyawaki, T. Tamegai, S. Awaji, H. Kito, S. Ishida, and Y. Yoshida, Supercond. Sci. Technol. **33**, 065001 (2020).

[5] M. Rotter, M. Tegel, and D. Johrendt, Phys. Rev. Lett. **101**, 107006 (2008).

[6] Z-C. Wang, C-Y. He, S-Q. Wu, Z-T. Tang, Y. Liu, A. Ablimit, C-M. Feng, and G-H. Cao, J. Am. Chem. Sco. **138**, 7856 (2016).

[7] Z. Wang, C. He, Z. Tang, S. Wu, G. Cao, Sci. China Mater. **60**, 83 (2017).

[8] Z.-C. Wang, C.-Y. He, S.-Q. Wu, Z.-T. Tang, Y. Liu, A. Ablimit, Q. Tao, C.-M. Feng, Z.-A. Xu, G.-H. Cao, J. Phys.: Condens. Matter **29**, 11LT01 (2017).

[9] Z.-C. Wang, C.-Y. He, S.-Q. Wu, Z.-T. Tang, Y. Liu, G.-H. Cao, Chem. Mater. **29**, 1805 (2017).

[10] S.-Q. Wu, Z.-C. Wang, C.-Y. He, Z.-T. Tang, Y. Liu, G.-H. Cao, Phys. Rev. Mater. **1**, 044804 (2017).

[11] X. Dong, K. Jin, D. Yuan, H. Zhou, J. Yuan, Y. Huang, W. Hua, J. Sun, P. Zheng, W. Hu, Y. Mao, M. Ma, G. Zhang, F. Zhou, and Z. Zhao, Phys. Rev. B **92**, 064515 (2015).

[12] S. Sun, S. Wang, R. Yu, and H. Lei, Phys. Rev. B **96**, 064512 (2017).

[13] G. Wang, Z. Wang, and X. Shi, EPL **116**, 37003 (2016).

[14] J. Ishida, S. Iimura, H. Hosono, Phys. Rev. B **96**, 174522 (2017).

[15] B. Wang, Z.-C. Wang, K. Ishigaki, K. Matsubayashi, T. Eto, J. Sun, J.-G. Cheng, G.-H. Cao, Y. Uwatoko, Phys. Rev. B **99**, 014501 (2019).

[16] M. Smidman, F. K. K. Kirschner, D. T. Adroja, A. D. Hillier, F. Lang, Z. C. Wang, G. H. Cao, and S. J. Blundell, Phys. Rev. B **97**, 060509(R) (2018).

[17] T. Wang, J. Chu, H. Jin, J. Feng, L. Wang, Y. Song, C. Zhang, X. Xu, W. Li, Z. Li, T. Hu, D. Jiang, W. Peng, X. Liu, and G. Mu, J. Phys. Chem. C **123**, 13925 (2019).

[18] Z.-C. Wang, Y. Liu, S.-Q. Wu, Y.-T. Shao, Z. Ren, and G.-H. Cao, Phys. Rev. B **99**, 144501 (2019).

[19] A. B. Yu, T. Wang, Y. F. Wu, Z. Huang, H. Xiao, G. Mu, and T. Hu, Phys. Rev. B 100, 144505 (2019).

[20] W. Hong, L. Song, B. Liu, Z. Li, Z. Zeng, Y. Li, D. Wu, Q. Sui, T. Xie, S. Danilkin, H. Ghosh, A. Ghosh, J. Hu, L. Zhao, X. Zhou, X. Qiu, S. Li, and H. Luo, arXiv:2005.06146v1 (2020).

[21] H. Q. Yuan, J. Singleton, F. F. Balakirev, S. A. Baily, G. F. Chen, J. L. Luo, and N. L. Wang, Nature (London) **457**, 565 (2009).

[22] M. M. Altarawneh, K. Collar, C. H. Mielke, N. Ni, S. L. Bud'ko, and P. C. Canfield, Phys. Rev. B **78**, 220505 (2008).

[23] Y. Nakajima, Y. Tsuchiya, T. Taen, T. Tamegai, S. Okayasu, and M. Sasase, Phys. Rev. B **80**, 012510 (2009).

[24] Y. Sun, Y. Tsuchiya, S. Pyon, T. Tamegai, C. Zhang, T. Ozaki, and Q. Li, Supercond. Sci. Technol. **28**, 015010 (2015).

[25] M. Nakajima, S. Ishida, T. Tanaka, K. Kihou, Y. Tomioka, T. Saito, C. H. Lee, H. Fukazawa, Y. Kohori, T. Kakeshita, A. Iyo, T. Ito, H. Eisaki, and S. Uchida, Sci. Rep. **4**, 5873 (2014).

[26] W. R. Meier, T. Kong, U. S. Kaluarachchi, V. Taufour, N. H. Jo, G. Drachuck, A. E. Böhmer, S. M. Saunders, A. Sapkota, A. Kreyssig, M. A. Tanatar, R. Prozorov, A. I. Goldman, Fedor F. Balakirev, Alex Gurevich, S. L. Bud'ko, and P. C. Canfield, Phys. Rev. B **94**, 064501 (2016).

[27] S. Pyon, A. Takahashi, I. Veshchunov, T. Tamegai, S. Ishida, A. Iyo, H. Eisaki, M. Imai, H. Abe, T. Terashima, and A. Ichinose, Phys. Rev. B **99**, 104506 (2019).

[28] A. F. Ioffe, and A. R. Regel, Prog. Semicond. 4, 237(1960).

[29] N. E. Husssey, K. Takenaka, and H. Takagi, Phil. Mag. **84**, 2847 (2004).

[30] Y. Liu, M. A. Tanatar, W. E. Straszheim, B. Jensen, K. W. Dennis, R. W. McCallum, V. G. Kogan, R. Prozorov, and T. A. Lograsso, Phys. Rev. B 89, 134504 (2014).

[31] N. R. Werthamer, E. Helfand, and P. C. Hohenberg, Phys. Rev. 147, 295 (1966).

[32] Y. Jia, P. Cheng, L. Fang, H. Luo, H. Yang, C. Ren, L. Shan, C. Gu, and H.-H. Wen, Appl. Phys. Lett. **93**, 032503 (2008).

[33] Z.-S. Wang, H.-Q Luo, C. Ren, and H.-H. Wen, Phys. Rev. B: Condens. Matter Mater. Phys. **78**, 140501(R) (2008).

[34] X. F. Wang, T. Wu, G. Wu, H. Chen, Y. L. Xie, J. J. Ying, Y. J. Yan, R. H. Liu, and X. H. Chen, Phys. Rev. Lett. **102**, 117005 (2009).

[35] M. A. Tanatar, N. Ni, G. D. Samolyuk, S. L. Bud'ko, P. C. Canfield, and R. Prozorov, Phys. Rev. B **79**, 134528 (2009).

[36] M. A. Tanar, M. S. Torikachvili, A. Thaler, S. L. Bud'ko, P. C. Canfield, and R. Prozorov, Phys. Rev. B **90**, 104518 (2014).

[37] Y. Nakamura, and S. Uchida, Phys. Rev. B **47**, 8369 (1993).

[38] K. Takenaka, K. Mizuhashi, H. Takagi, and S. Uchida, Phys. Rev. B **50**, 6534 (1994).

[39] T. Watanabe, T. Fujii, and A. Matsuda, Phys. Rev. Lett. **79**, 2113 (1997).

[40] M. A. Tanatar, N. Ni, A. Thaler, S. L. Bud'ko, P. C. Canfield, and R. Prozorov, Phys. Rev. B **82**, 134528 (2010).

[41] M. A. Tanatar, N. Ni, A. Thaler, S. L. Bud'ko, P. C. Canfield, and R. Prozorov, Phys. Rev. B **84**, 014519 (2011).

[42] M. A. Tanatar, E. C. Blomberg, H. Kim, K. Cho, W. E. Straszheim, B. Shen, H.-H. Wen, and R. Prozorov, arXiv:1106.0533 (2011).

[43] Y. Nakajima, T. Taen, and T. Tamegai, J. Phys. Soc. Jpn. **78**, 023702 (2009).

[44] K. Ohgushi, and Y. Kiuchi, Phys. Rev. B **85**, 064522 (2012).

[45] Y. Nakajima, Y. Tsuchiya, T. Taen, T. Tamegai, S. Okayasu, and M. Sasase, Phys. Rev. B **80**, 012510 (2009).

[46] C. P. Bean, Rev. Mod. Phys. 36, 31 (1964).

[47] E. M. Gyorgy, R. B. van Dover, K. A. Jackson, L. F. Schneemeyer, and J. V. Waszczak, Appl. Phys. Lett. 55, 283 (1989).

[48] A. Park, S. Pyon, K. Ohara, N. Ito, and T. Tamegai, Phys. Rev. B **97**, 064516 (2018).

[49] T. Tamegai, T. Taen, H. Yagyuda, Y. Tsuchiya, S. Mohan, T. Taniguchi, Y. Nakajima, S. Okayasu, M. Sasase, H. Kitamura, T. Murakami, T. Kambara, and Y. Kanai, Supercond. Sci. Technol. **25**, 084008 (2012).

[50] A. Park, I. Veshchunov, A. Mine, S. Pyon, T. Tamegai, and H. Kitamura, accepted in Phys. Rev. B.

[51] T. Taen, F. Ohtake, S. Pyon, T. Tamegai, and H. Kitamura, Supercond. Sci. Technol. **28**, 085003 (2015).

[52] F. Ohtake, T. Taen, S. Pyon, T. Tamegai, S. Okayasu, T. Kambara, and H. Kitamura, Physica C **518**, 47 (2015).

[53] Y. Sun, T. Taen, Y. Tsuchiya, Q. Ding, S. Pyon, Z. Shi, and T. Tamegai, Appl. Phys. Express **6**, 043101 (2013).

[54] S. J. Singh, M. Bristow, W. R. Meier, P. Taylor, S. J. Blundell, P. C. Canfield, and A. I. Coldea, Phys. Rev. Mater. **2**, 074802 (2018).

[55] S. Ishida, A. Iyo, H. Ogino, H. Eisaki, N. Takeshita, K. Kawashima, K. Yanagisawa, Y. Kobayashi, K. Kimoto, H. Abe, M. Imai, J. Shimoyama, and M. Eisterer, npj Quantum Materials **4**, 27 (2019).

[56] N. Ito, S. Pyon, T. Kambara, A. Yoshida, S. Okayasu, A. Ichinose, and T. Tamegai, J. Phys.: Conf. Series **1054**, 012020 (2018).

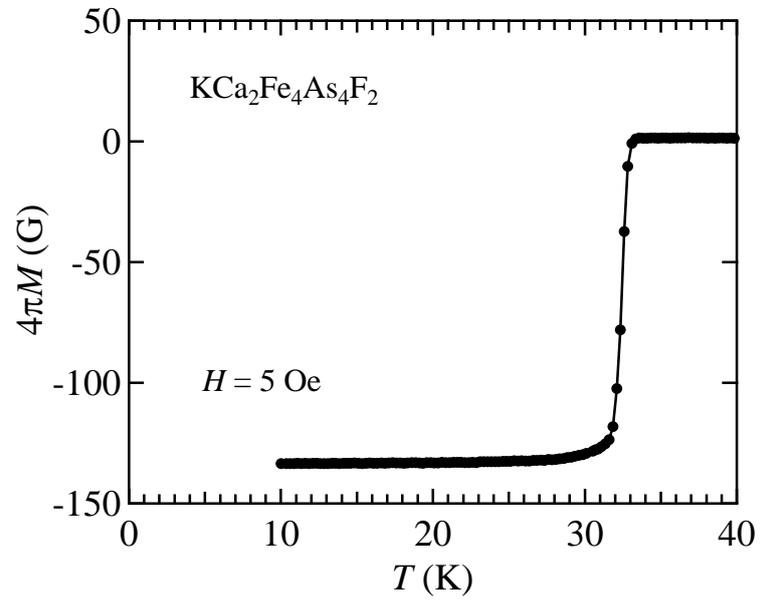

Fig. 1. Temperature dependence of the zero-field-cooled magnetization at $H = 5$ Oe along the $c$-axis in $KCa_2Fe_4As_4F_2$.

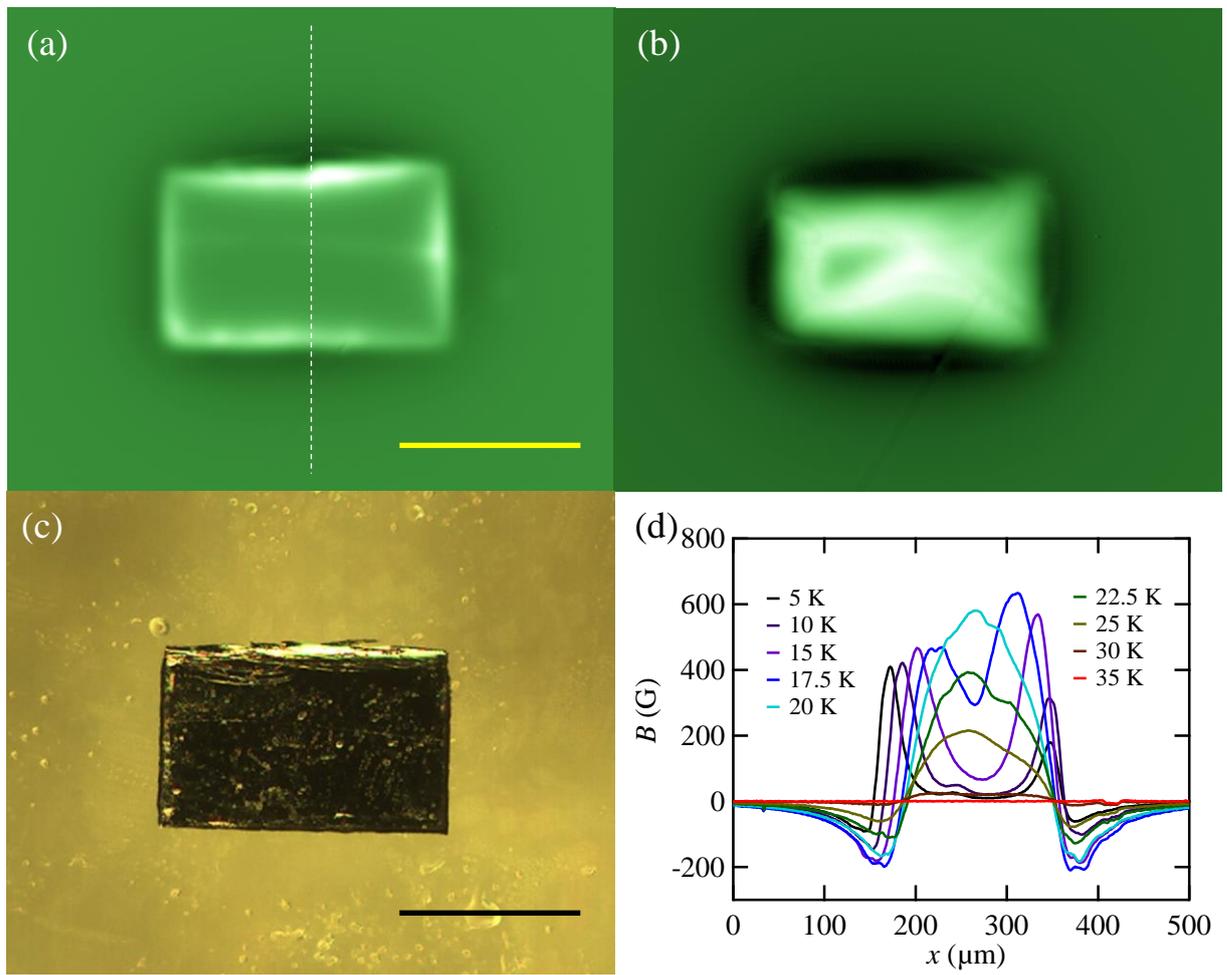

Fig. 2. Differential MO images of KCa$_2$Fe$_4$As$_4$F$_2$ in the remanent state at (a) 5 K and (b) 20 K after cycling the field up to 1.6 kOe for 0.2 s. (c) An optical micrograph of KCa$_2$Fe$_4$As$_4$F$_2$. (d) Local magnetic induction profiles at different temperatures taken along the broken line in (a). Both the yellow bar in (a) and black bar in (c) correspond to 200 μm.

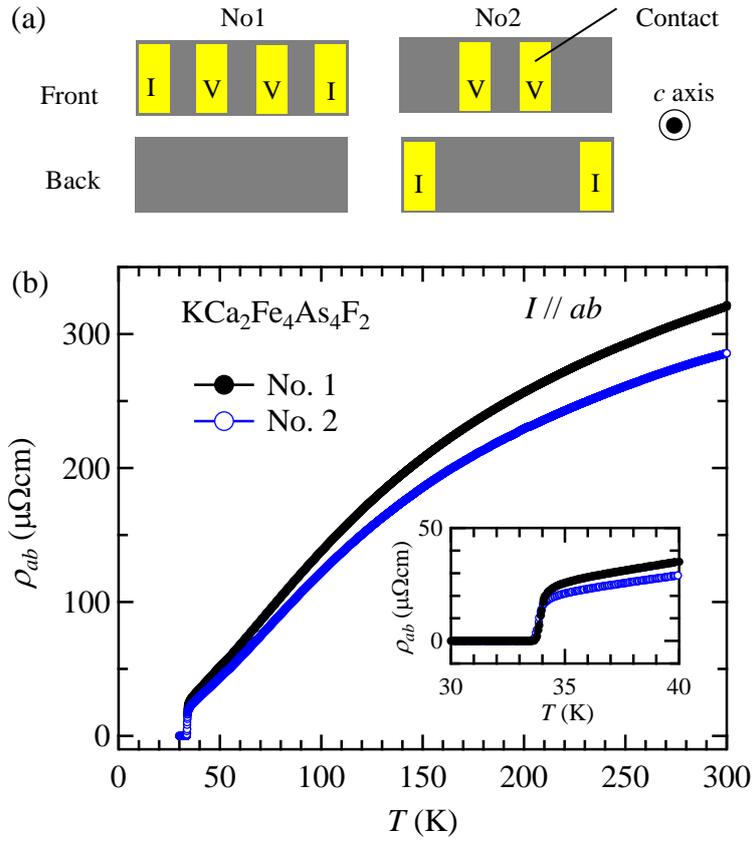

Fig. 3 (a) Schematic drawings of the two samples with electrical contacts for measurements of in-plane electrical resistivity ($\rho_{ab}$). (b) Temperature dependence of $\rho_{ab}$ in KCa$_2$Fe$_4$As$_4$F$_2$ sample No. 1 and No. 2 measured in a temperature range of 30-300 K. The inset in (b) shows $\rho_{ab}-T$ near $T_c$.

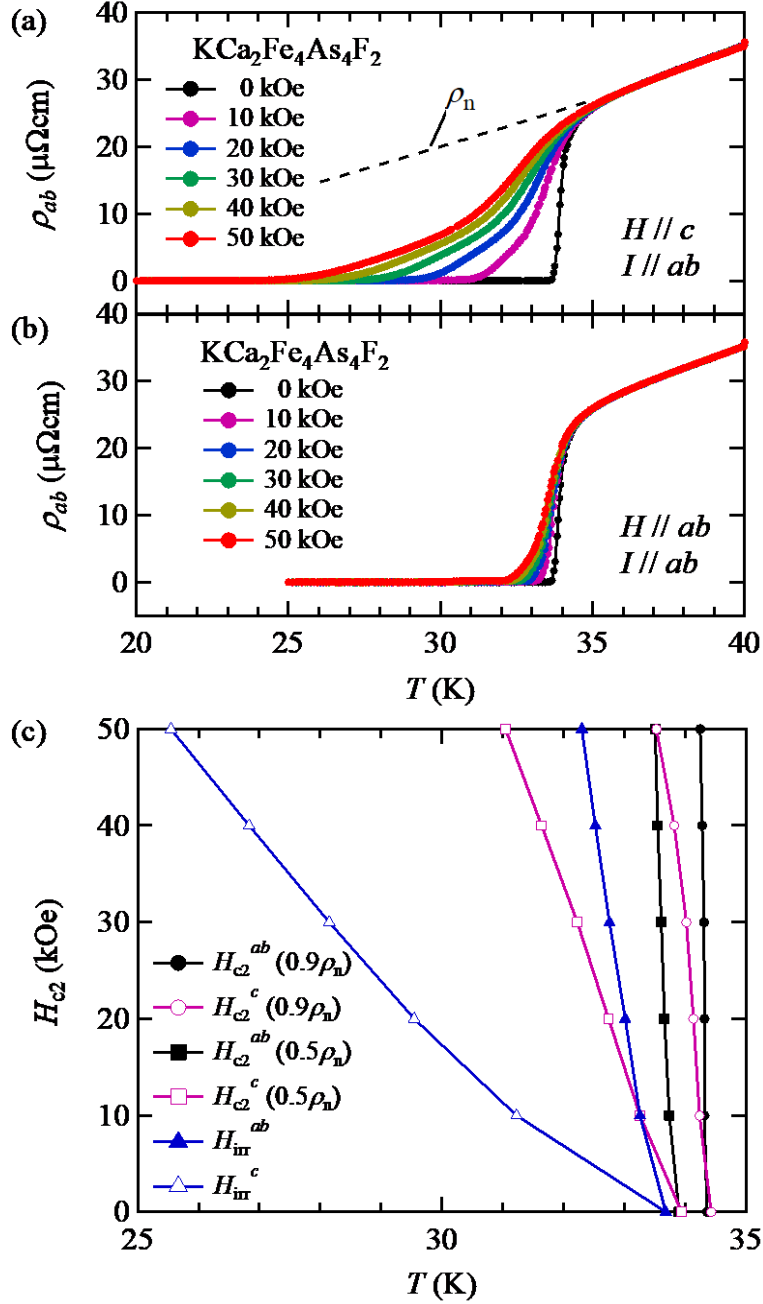

Fig. 4. Temperature dependence of in-plane electrical resistivity ($\rho_{ab}$) in KCa$_2$Fe$_4$As$_4$F$_2$ (No.1) below 40 K under various magnetic fields parallel to the (a) $c$-axis and (b) $ab$-plane. (c) Anisotropic $H_{c2}$ and $H_{irr}$ evaluated from temperature-dependent resistivity in (a) and (b). $H_{c2}$ is defined by two different criteria of 0.9$\rho_n$ and 0.5 $\rho_n$. Here, $\rho_n$ is the normal state resistivity estimated from the extrapolation of the resistivity using the power-law form with $n$ = 1.3 as shown in the broken line in (a). $H_{irr}$ is defined by the criteria of $\rho_{ab}$ ~ 0.5 $\mu\Omega$cm.

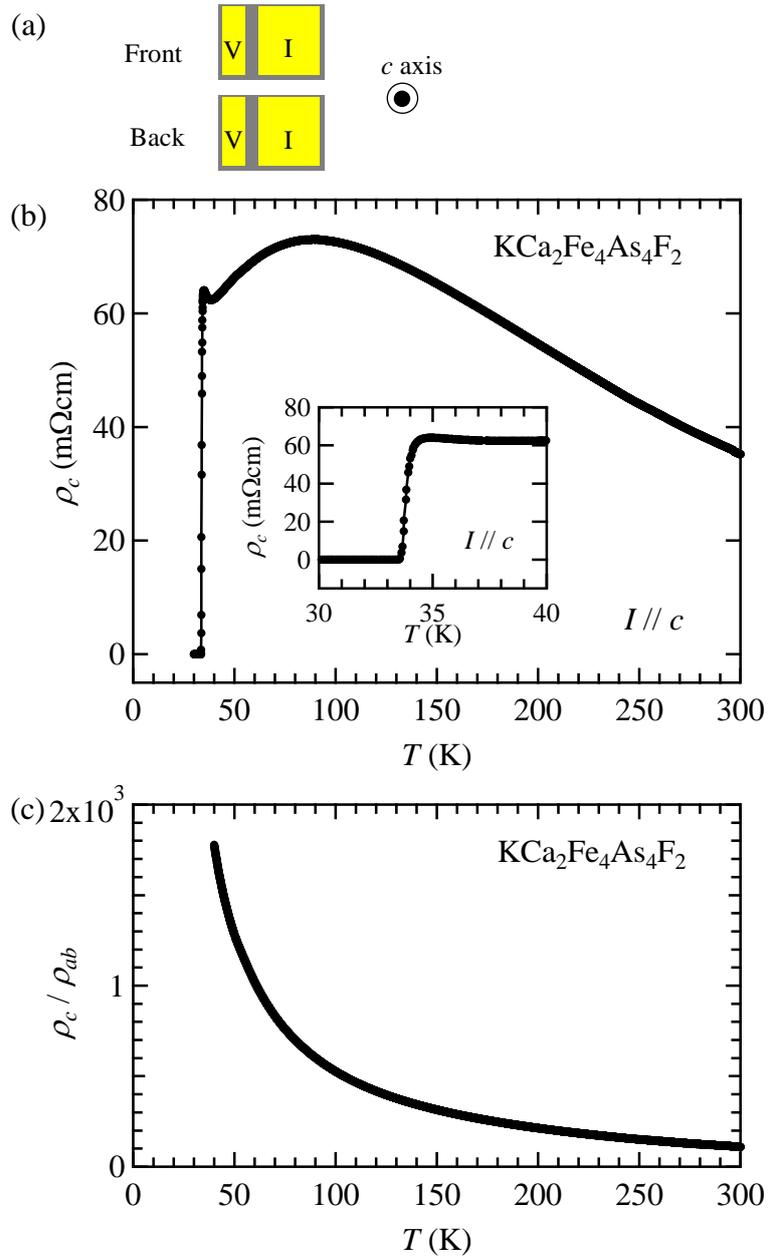

Fig. 5. (a) Schematic drawings of the electrical contacts for the measurement of out-plane electrical resistivity ($\rho_c$). (b) Temperature dependence of $\rho_c$ in $KCa_2Fe_4As_4F_2$ in a temperature range of 30-300 K under zero magnetic field. The inset in (b) shows $\rho_c$-$T$ near $T_c$. (c) Temperature dependence of the anisotropy of the resistivity, $\rho_c / \rho_{ab}$. $\rho_{ab}$ of sample No.1 shown in Fig. 3 (b) is used for the calculation.

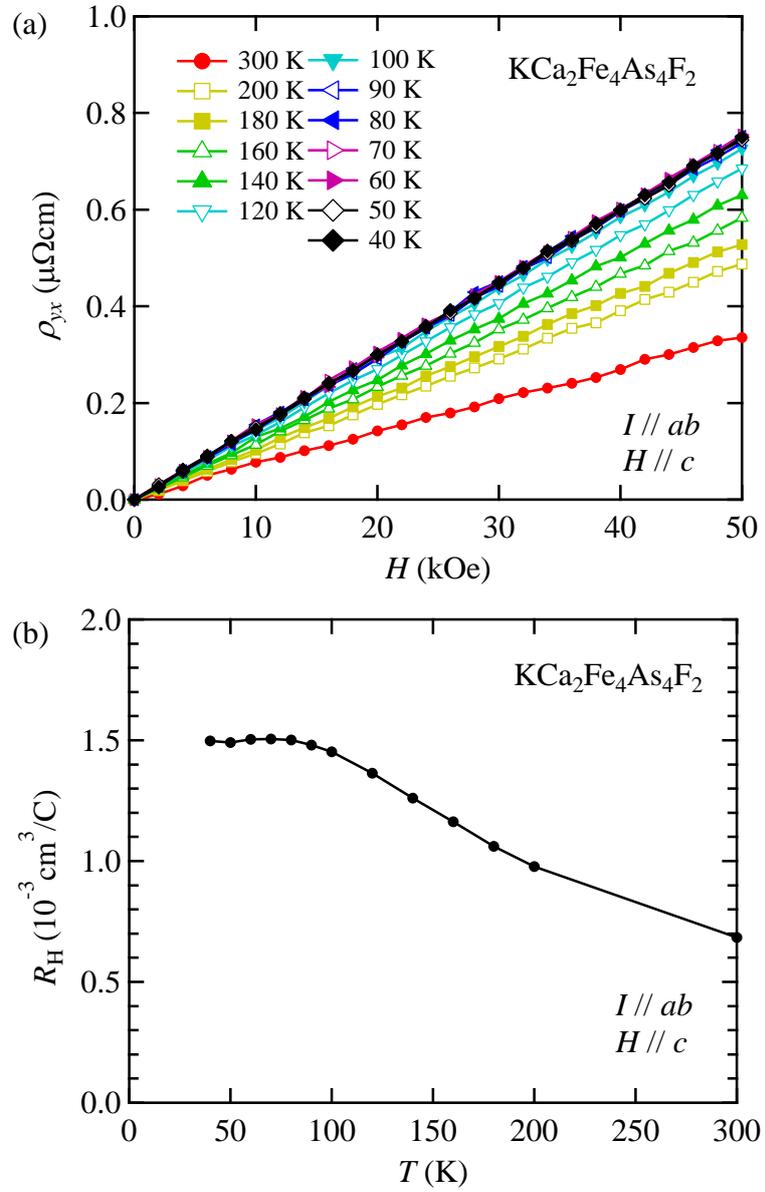

Fig. 6. (a) Hall resistivity $\rho_{yx}$ as a function of field at various temperatures in $KCa_2Fe_4As_4F_2$. (b) Temperature dependence of the Hall coefficient $R_H$ in $KCa_2Fe_4As_4F_2$.

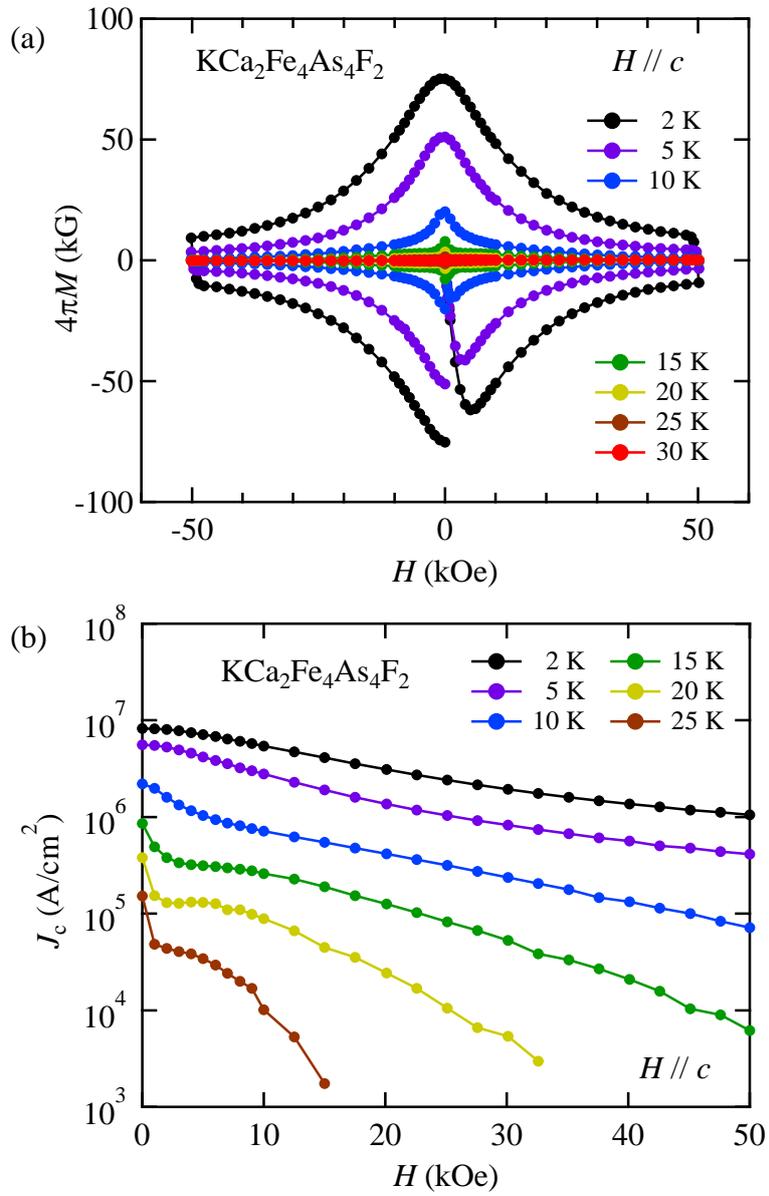

Fig. 7. (a) Magnetic field dependence of magnetization in $KCa_2Fe_4As_4F_2$ at various temperatures for $H//c$-axis. (b) Magnetic field dependence of in-plane $J_c$ evaluated using the data shown in (a).

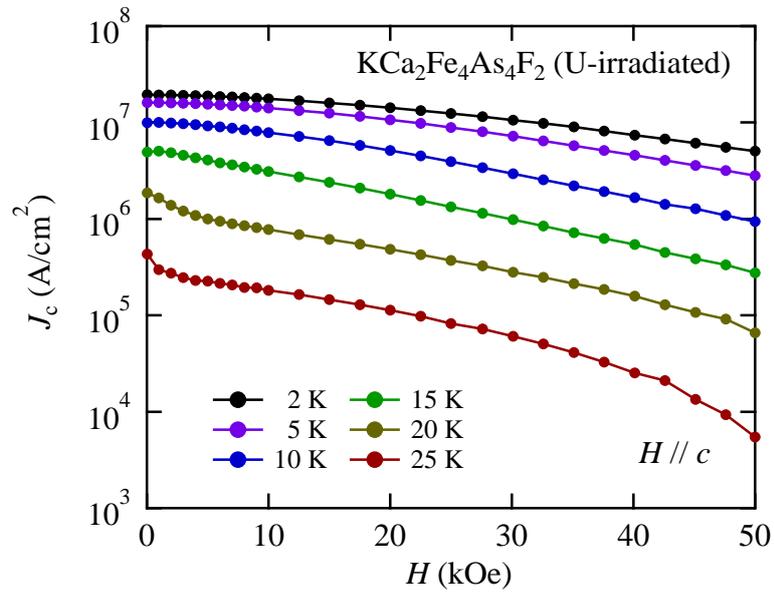

Fig. 8. Magnetic field dependence of magnetic $J_c$ at various temperature in 2.6 GeV U-irradiated ($B_\Phi$ = 40 kOe) $KCa_2Fe_4As_4F_2$.

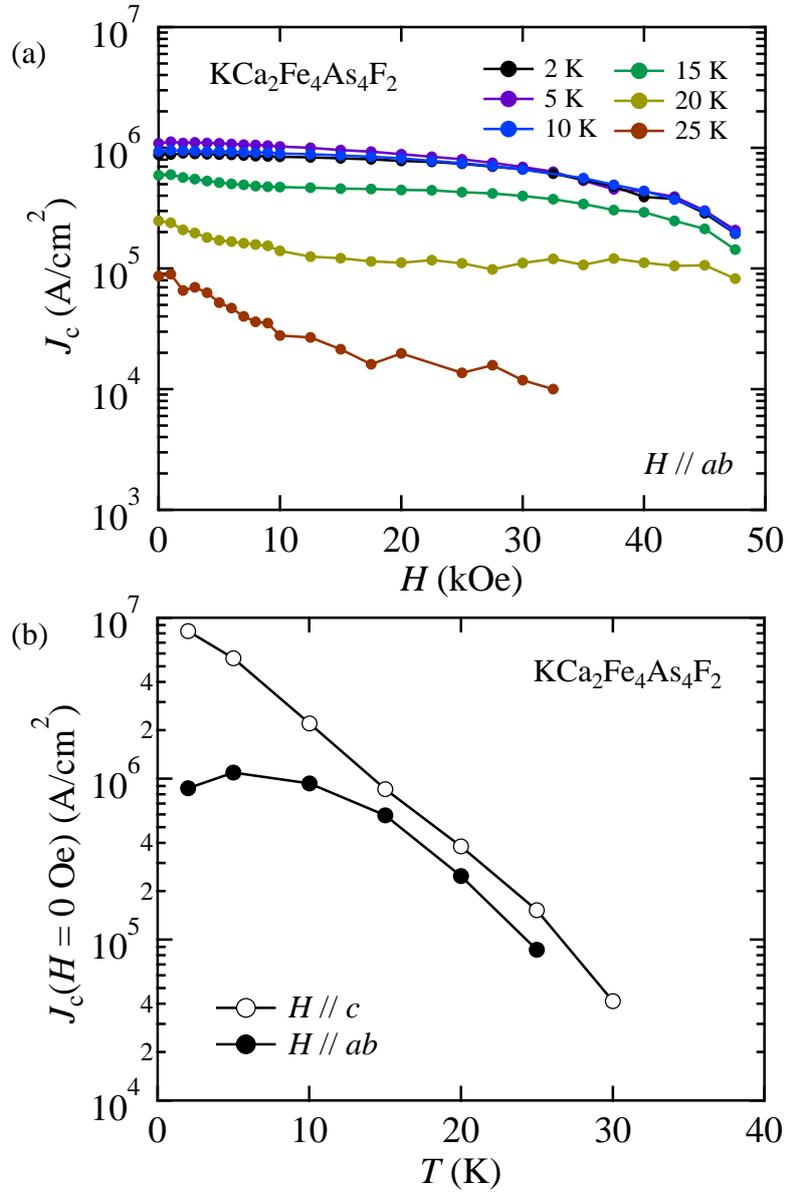

Fig. 9. (a) Magnetic field dependence of magnetic $J_c$ in $KCa_2Fe_4As_4F_2$ at various temperatures for $H//ab$-plane. (b) Temperature dependence of magnetic $J_c$ under the self-field in $KCa_2Fe_4As_4F_2$ for $H//c$-axis and $H//ab$-plane.

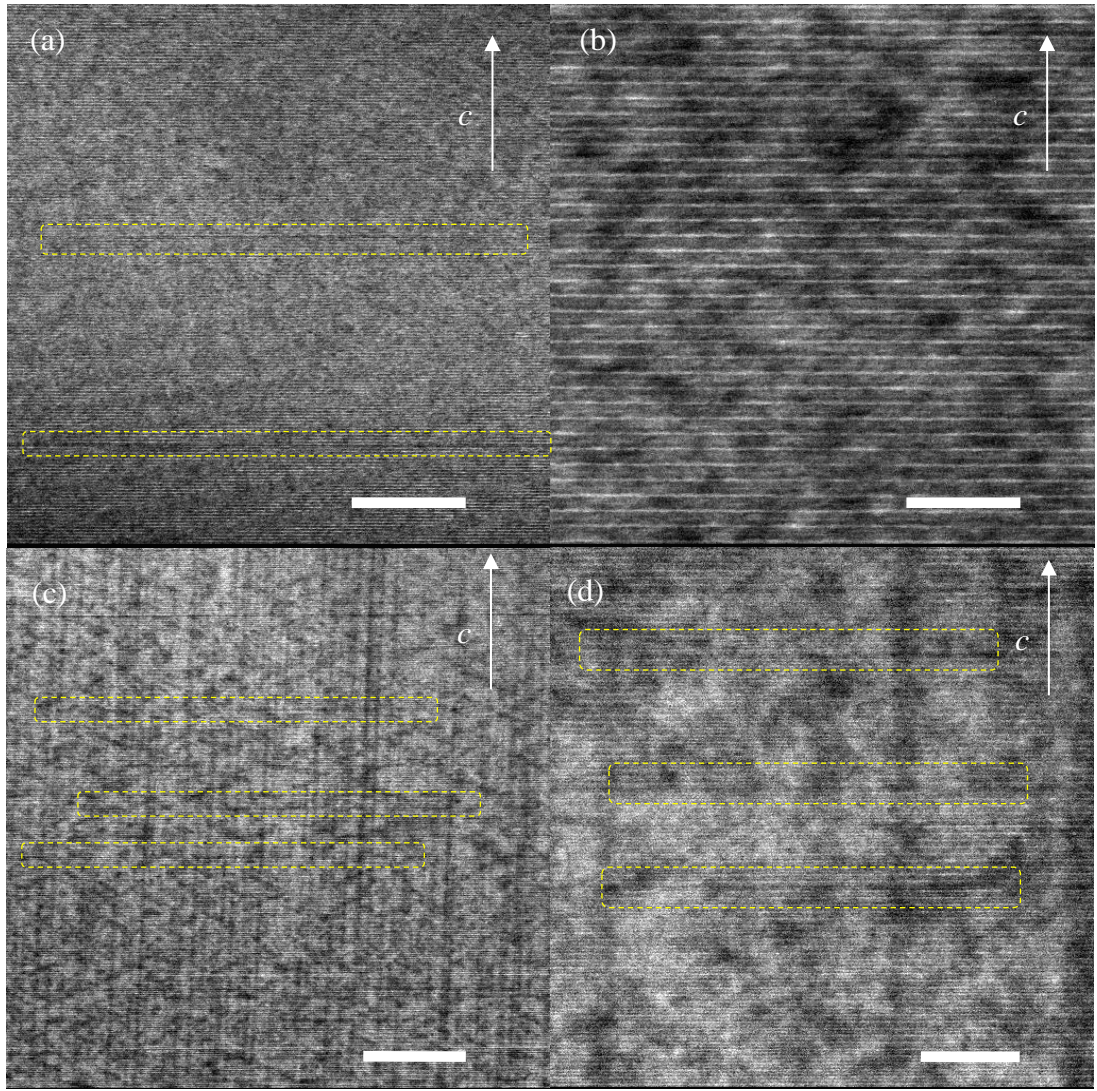

Fig. 10. STEM images of (a), (b) the pristine and (c), (d) 2.6 GeV U-irradiated $KCa_2Fe_4As_4F_2$ for an electron beam injected along the *a* axis. Solid lines in (a)-(d) correspond to 100, 10, 100, and 20 nm, respectively. Yellow broken squares in (a), (c), (d) emphasize the location of horizontal black line-like defects.